\documentclass[conference,10pt]{IEEEtran}
\usepackage[noadjust]{cite}
\usepackage{cite}
\usepackage{amsmath,amssymb,amsfonts}
\usepackage{algorithmic}
\usepackage{graphicx}
\usepackage{textcomp}
\usepackage{xcolor}
\usepackage{pgfplotstable}
\usepackage{dirtytalk}  % quotation marks
\usepackage{pgfplots}
\usepackage{hyperref}
\usepackage{subcaption}
\usepackage{ifthen}
\usetikzlibrary{shapes.arrows}

\usepackage[frozencache=true,cachedir=minted-cache]{minted}  % use without sheel-escape aka for arxiv
\makeatletter
\newcommand\notsotiny{\@setfontsize\notsotiny\@vipt\@viipt}
\makeatother
\usepackage{enumitem}
\usepackage{amsmath} % for cases environment and \text command
\usepackage{xcolor}
\usepackage[skip=3pt]{caption}
\usepackage{booktabs}
\usetikzlibrary{automata, positioning, arrows, matrix, calc,3d}
\usetikzlibrary{intersections}

\definecolor{GoogleGreen}{RGB}{15,200,88}
\definecolor{GoogleRed}{RGB}{250,68,55}
\definecolor{GoogleYellow}{RGB}{244,200,0}
\definecolor{green2}{RGB}{100,182,83}
\definecolor{dkgreen}{rgb}{0,0.6,0}
\definecolor{gray}{rgb}{0.5,0.5,0.5}
\definecolor{mauve}{rgb}{0.58,0,0.82}
\definecolor{darkblue}{rgb}{0.0,0.0,0.6}
\definecolor{cyan}{rgb}{0.0,0.6,0.6}
\definecolor{mBlue}{HTML}{4285f4}
\definecolor{mRed}{HTML}{ea4335}

\definecolor{blue1}{HTML}{BFBFFF}
\definecolor{blue2}{HTML}{A3A3FF}
\definecolor{blue3}{HTML}{7879FF}
\definecolor{blue4}{HTML}{4949FF}
\definecolor{blue9}{HTML}{7B9FF2}

\definecolor{red1}{HTML}{F6BDC0}
\definecolor{red2}{HTML}{F1959B}
\definecolor{red3}{HTML}{F07470}
\definecolor{red4}{HTML}{EA4C46}

\definecolor{mGreen}{HTML}{34a853}
\definecolor{mYellow}{HTML}{fbbc04}
\definecolor{mLightBlue}{HTML}{6d9eeb}
\definecolor{mLightRed}{HTML}{e06666}
\definecolor{mLightGreen}{HTML}{93c47d}
\definecolor{mLightYellow}{HTML}{ffd966}
\definecolor{archtBlue}{HTML}{9fc5e8}
\definecolor{archtTeal}{HTML}{a4d2d7}
\definecolor{archtYellow}{HTML}{ffe599}
\definecolor{archtOrange}{HTML}{f9cb9c}
\definecolor{archtCetus}{HTML}{ea9999}
\definecolor{archtOther}{HTML}{dd7e6b}
\definecolor{archtPurple}{HTML}{b4a7d6}
\definecolor{archtGray}{HTML}{eeeeee}
\definecolor{archtGreen}{HTML}{b6d7a8}
\definecolor{archtRed}{HTML}{ea9999}

\newcommand{\backus}[0]{\textsc{ScalSALE}}
\newcommand{\LULESH}[0]{\textsc{Lulesh}}
 
\def\BibTeX{{\rm B\kern-.05em{\sc i\kern-.025em b}\kern-.08em
    T\kern-.1667em\lower.7ex\hbox{E}\kern-.125emX}}
\pgfplotsset{compat=1.4}
\DeclareRobustCommand*{\IEEEauthorrefmark}[1]{%
  \raisebox{0pt}[0pt][0pt]{\textsuperscript{\footnotesize\ensuremath{#1}}}}
  \usepackage[misc]{ifsym}
\newcommand{\corrAuthor}{$^{\textrm{\Letter}}$}
\ifCLASSINFOpdf
\else
\fi
\DeclareRobustCommand*{\IEEEauthorrefmark}[1]{%
  \raisebox{0pt}[0pt][0pt]{\textsuperscript{\footnotesize\ensuremath{#1}}}}
  \usepackage[misc]{ifsym}

\begin{document}
\bstctlcite{IEEEexample:BSTcontrol}

\title{\backus{}: Scalable SALE Benchmark Framework for Supercomputers}

\author{\IEEEauthorblockN{Re'em Harel\IEEEauthorrefmark{1,2,3},
Matan Rusanovsky\IEEEauthorrefmark{2,3},
Ron Wagner\IEEEauthorrefmark{4,6},
Harel Levin\IEEEauthorrefmark{2,3} and
Gal Oren\IEEEauthorrefmark{2,5\hspace{0.1cm}$\corrAuthor$}}\\
\IEEEauthorblockA{\IEEEauthorrefmark{1}Department of Computer Science, Ben-Gurion University of the Negev, Israel}
\IEEEauthorblockA{\IEEEauthorrefmark{2}Scientific Computing Center, Nuclear Research Center – Negev, Israel}
\IEEEauthorblockA{\IEEEauthorrefmark{3}Department of Physics, Nuclear Research Center – Negev, Israel}
\IEEEauthorblockA{\IEEEauthorrefmark{4}Israel Atomic Energy Commission}
\IEEEauthorblockA{\IEEEauthorrefmark{5}Department of Computer Science, Technion – Israel Institute of Technology, Israel}
\IEEEauthorblockA{\IEEEauthorrefmark{6}School of Physics and Astronomy, Tel Aviv University, Israel}
{\tt\small reemha@bgu.ac.il, matanr@nrcn.org.il, ronwagner@mail.tau.ac.il,  harellevin@nrcn.org.il,}\\ {\tt\small galoren@cs.technion.ac.il}
}

% \author{
% \IEEEauthorblockN{
% Re'em Harel\IEEEauthorrefmark{1,2,3},
% Matan Rusanovsky\IEEEauthorrefmark{2,3},
% Ron Wagner\IEEEauthorrefmark{4,6},
% % Harel Levin\IEEEauthorrefmark{2,3} and
% Gal Oren\IEEEauthorrefmark{2,5\hspace{0.1cm}\corrAuthor}}\\

% \IEEEauthorblockA{\IEEEauthorrefmark{1}Department of Computer Science, Ben-Gurion University of the Negev, Israel}
% \IEEEauthorblockA{\IEEEauthorrefmark{2}Scientific Computing Center, Nuclear Research Center – Negev, Israel}
% \IEEEauthorblockA{\IEEEauthorrefmark{3}Department of Physics, Nuclear Research Center – Negev, Israel}
% \IEEEauthorblockA{\IEEEauthorrefmark{4}Israel Atomic Energy Commission}
% \IEEEauthorblockA{\IEEEauthorrefmark{5}Department of Computer Science, Technion – Israel Institute of Technology, Israel}
% \IEEEauthorblockA{\IEEEauthorrefmark{6}School of Physics and Astronomy, Tel Aviv University, Israel}
% }
%{\tt\small reemha@bgu.ac.il, matanr@nrcn.org.il, ronwagner@mail.tau.ac.il,  harellevin@nrcn.org.il,
%galoren@cs.technion.ac.il}
% \\[-1.5ex]
%}

\IEEEtitleabstractindextext{

\begin{abstract}
Supercomputers worldwide provide the necessary infrastructure for groundbreaking research. As the demand for enhancing their performance constantly increases, new systems are built worldwide to supply this demand at an increased pace, volume, and power. However, most supercomputers are not designed equally due to different desired figure of merit, which is derived from the computational bounds of the targeted scientific applications' portfolio. In turn, the design of such computers becomes a budget-constrained optimization process that strives to achieve the best performances possible in a multi-parameters search space, including heterogeneous hardware, middleware, and software. Therefore, verifying and evaluating whether a supercomputer can achieve its desired goal becomes a tedious and complex task. For this purpose, many full, mini, proxy, and benchmark applications have been introduced in the attempt to represent scientific applications partially. Nevertheless, as these benchmarks are hard to expand, update technologically, and most importantly, are over-simplified compared to scientific applications that tend to couple multiple scientific domains, they fail to represent the true scaling capabilities. We suggest a new physical scalable benchmark framework, namely \backus{}, based on the well-known SALE scheme. \backus{}'s main goal is to provide a simple, flexible, scalable infrastructure that can be easily expanded to include multi-physical schemes while maintaining scalable and efficient execution times. By expanding \backus{}, the gap between the over-simplified benchmarks and scientific applications can be bridged. To achieve this goal, \backus{} is implemented in Modern Fortran with simple object-oriented design patterns and supported by transparent MPI-3 blocking and non-blocking communication that allows such a scalable framework. In this manner, \backus{} easily achieves almost identical performances in similar settings to common original benchmark suites of its kind while allowing for expansion and support for hardware and software modernization. The usage of \backus{} is demonstrated with the multi-bounded representative Sedov-Taylor blast wave problem and compared to the well-known \LULESH{} benchmark using strong and weak scaling tests. \backus{} is executed and evaluated with both rezoning options -- Lagrangian and Eulerian, to demonstrate how the benchmark can be expanded to solve additional problems while maintaining scalable execution times. The framework, as well as the results and extensions, are available at: \textcolor{blue}{\url{https://github.com/Scientific-Computing-Lab-NRCN/ScalSALE}}.

\end{abstract}

\begin{IEEEkeywords}
LULESH, Modern Fortran, MPI-3, HPC, Parallel OOP, Scalabale Framework, Benchmark, oneAPI
\end{IEEEkeywords}}

\maketitle
\IEEEdisplaynontitleabstractindextext

\IEEEpeerreviewmaketitle
% \begin{abstract}
% %Top Supercomputers worldwide (still, and mostly) rely on Fortran code for peak performance computation. However, most currently available scaling benchmarks for said systems are either coded differently (C, C++), do not correspond to an actual computation (HPL), or fail to scale. We present a novel Modern Fortran Object-Oriented benchmarks framework that can be easily expanded to new computation scenarios while transparently applying MPI-3 advanced distribution for both weak and strong scaling. We demonstrate these features by implementation and comparison to the known LULESH benchmark.
% Evaluating and assessing the performance of supercomputers is a crucial step. The most common way to evaluate is by measuring benchmarks...
% \end{abstract}

% \begin{IEEEkeywords}
% Modern Fortran, Benchmark, MPI, HPC, Scientific Code
% \end{IEEEkeywords}
% \settopmatter{printfolios=true}
\vspace{-4ex}
\section{Introduction}

\subsection{Objectives and Designs of Supercomputers}
The combination of high-performance computing (high resolution and fidelity) and storage (large outputs to be analyzed) in supercomputers provide the essential infrastructure for groundbreaking research, simulation modeling, and data analysis in various scientific domains~\cite{wilson1989grand}.

Most, if not all, scientific applications' performance tends to be bounded by a computer component. For example, CPU-bound applications are limited by the processing speed of the CPUs, thus, they can be sped up by adding more cores to the machine; I/O-bound applications are limited by the amount of time the application spends waiting for I/O operations; memory-bound applications are limited by the memory access speed and the amount of available memory; latency-bound applications are limited by the time it takes to retrieve data; bandwidth-bound are limited by the amount of data that needs to be retrieved (and transferred). 

It is common to classify scientific applications into several algorithmic classes or kernels that can be associated with a specific bound~\cite{web:Aurora-presentation}. For example, dense linear algebra~\cite{bientinesi2005science} and structured grids with high flop/s rate are classified as compute-bound; sparse linear algebra and particle methods~\cite{particlemethod} with high-performance memory systems are classified as memory-bound; FFTs~\cite{brigham1988fast} with high bisection bandwidth is classified as bandwidth-bound; unstructured grid or adaptive mesh refinement (AMR)~\cite{amr} with a low latency network is classified as latency-bound; and data-intensive computations with storage and network access can be classified as memory and I/O-bound. Nevertheless, most applications are associated with multiple algorithmic classes, thus making them multi-bound.

% Generally, each scientific domain is studied via numerous scientific applications. In the core of these scientific applications lies several algorithm classes or kernels that can be associated with a specific bound~\cite{web:Aurora-presentation}. 

As the demand for performance enhancement of these algorithmic classes --- and in turn, the associated scientific applications --- constantly increases, new hardware architectures and designs have emerged over the years. For example, the introduction of Infiniband~\cite{infiniband} --- a high bandwidth and low-latency network, improved the performance of scientific applications that use unstructured or adaptive mesh refinement (AMR), such as in fusion~\cite{fusionmethods}, climate~\cite{climate} and astrophysics~\cite{kitchin2008astrophysical}. In the same way, general-purposed GPUs (GPGPUs)~\cite{gpgpus} reduced the execution time of dense linear algebra and FFTs calculations common in material science, chemistry, astrophysics, and deep learning applications. In addition, the introduction of high bandwidth memory (HBM)~\cite{jun2017hbm} and DDR5 enhanced memory-bound kernels such as particle methods. 
Moreover, data-intensive calculations were boosted with the introduction of burst buffer storage~\cite{bbs}. Thus, it can be claimed that the targeted applications' work fashion drives supercomputers' designs and architectures.

Due to the wide variety of scientific applications and their corresponding optimal hardware, designing a supercomputer has become much more complex and diverse compared to previous generations~\cite{khan2021analysis} (as can also be seen in the Top500 list~\cite{top500}), especially if it is designed under a budget. For example, the Fugaku~\cite{monroe2020fugaku} supercomputer, which holds the first spot in the HPCG list~\cite{dongarra2016high}, contains only CPUs and a low-latency network which makes it optimal for tightly-coupled simulations and modeling but not for artificial intelligence applications~\cite{stevens2019aurora}. On the contrary, Frontier~\cite{schneider2022exascale}, the first exascale supercomputer, is heterogeneous --- containing both CPUs and GPUs. More so, Facebook's supercomputer~\cite{fbsupercomputer} is intended for AI modeling exclusively. Thus, it contains almost only GPUs. Furthermore, the new HPC as a service (HPCaaS) allows the formation of large-scale cloud clusters with low network capabilities (high latency)~\cite{walkup2022best, gupta2011evaluation}, best for embarrassingly parallel and loosely coupled computations (with minor to moderate communication operations)~\cite{ali2013outlook}. Such a system is suited for deep learning applications and small-scale traditional simulations and modeling. %A unique case of a supercomputer is the cloud as a high-performance computing service. %It contains many GPGPUs, CPUs, and large storage units but has high-latency networks~\cite{walkup2022best, gupta2011evaluation}. Thus, suited for deep learning applications and small-scale traditional simulations and modeling. 

\subsection{Assessment and Evaluation of Supercomputers}
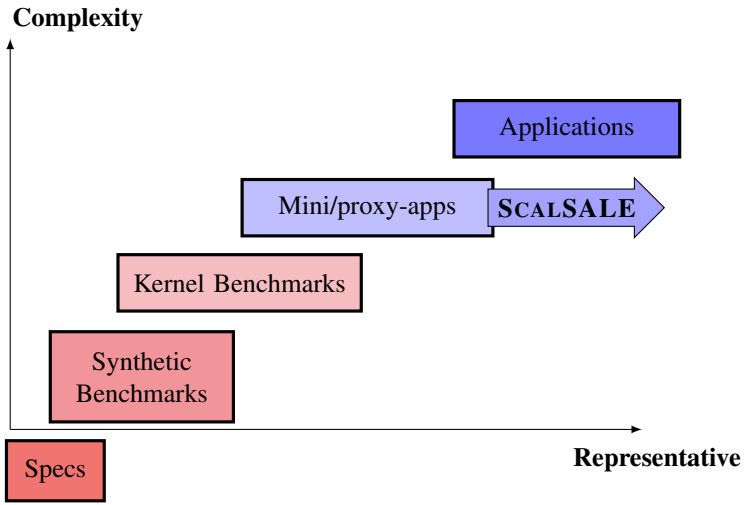
\begin{figure}
    \tikzstyle{rect} = [rectangle, draw, text centered, very thick, minimum width =2cm, minimum height=0.75cm]

\begin{tikzpicture}[node distance = 1.5cm, auto]

% \node[rect, fill=GoogleYellow,text width=1.3cm,align=center ] (statistics) at (1.1,0.75) {Basic \linebreak Statistics};

\node[rect, fill=red3,minimum height=0.8cm,minimum width=1.3cm] (skeleton) at (1.2, 0.25) {Specs};

\node[rect, fill=red2,minimum height=1.2cm,text width=2.2cm,align=center] (skeleton) at (2.35, 1.5) {Synthetic \linebreak Benchmarks};

\node[rect, fill=red1,text width=3cm,align=center, minimum height=0.75cm] (skeleton) at (3.65, 2.75) {Kernel Benchmarks};

\node[rect, fill=blue1, text width=3.1cm,align=center, minimum height=0.75cm] (skeleton) at (5.35, 3.75) {Mini/proxy-apps};

 \node[rect, fill=blue3, minimum width =3cm] (skeleton) at (8, 4.8) {Applications};
 
        \node[single arrow, draw,fill=blue2,
  minimum height=1cm, single arrow head extend=1.5mm]   at (8,3.75) {\textbf{\backus{}}};
    %  \node [single arrow,draw,minimum height=0.7cm, text width=1cm, single arrow head extend=0.8cm,anchor=west]  at (8.2, 3.75) {\textbf{\backus{}}};

%  \node[rectangle, dashed, minimum width =1cm] (skeleton) at (8.2, 3.75) {\textbf{\backus{}}};

% \draw[-latex, dashed, very thick] (6.5,3.75) -- (7.2, 3.75);
% \draw[-latex, dashed, very thick] (7.84,4.6) -- (7.84, 3.9);

    \draw [-latex](0.6,0.8) -- (0.6,6);
\node[inner sep=0pt] (whitehead) at (1.5,6.25)    {\textbf{Complexity}};

    \draw [-latex](0.6,0.8) -- (9,0.8);

\node[inner sep=0pt] (whitehead) at (9.2,0.4)    {\textbf{Representative}};

\end{tikzpicture}
\caption{The steps taken in order to evaluate and assess supercomputers as a function of their complexity and representation of scientific applications --- based on~\cite{lascu2009hpc}. \backus{} extends the mini/proxy apps, such that it will maintain the same complexity as most mini-apps while overlapping and representing scientific applications.}
    \label{fig:pyramid_evaluation}
        \vspace{-4ex}

\end{figure}

Over the years, the goals of supercomputers have become diverse --- supporting a wide portfolio of scientific applications --- and as such, the design and architectures are also varied. Thus, the crucial step of assessing, evaluating, and comparing supercomputers has become hard and complex. 

To cope with this hard task, the evaluation is usually broken down into different steps, associated with its complexity, and representation of the entire app as a whole --- from the most simple test to executing a whole application (an illustration of these steps can be seen in \autoref{fig:pyramid_evaluation}):

\begin{enumerate}[itemindent=-4pt,topsep=0pt,itemsep=-1ex,partopsep=0ex,parsep=1ex]
    \item Basic parameters and specifications --- include: vendor (AMD, Intel, Cray), efficiency, Rpeak, architecture (SMP, SIMD), operating system (RHEL, SUSE, Ubuntu), interconnect family (Ethernet, Infiniband), number of cores, accelerators family, application area (academic, industry), and more. These parameters provide a general overview of the machine and its composition. As this step is not complex at all, it is represented in \autoref{fig:pyramid_evaluation} below the x-axis.
    
    \item Simple Synthetic Benchmarks --- performed to test and validate a single-system component's basic functionality and behavior. Thus, they are less representative yet, easy to maintain and highly scalable. For example, IOR~\cite{ior} (random I/O operations), ALCF~\cite{morozov2012alcf} or MPIGraph~\cite{moody2007mpigraph} (basic MPI operations) and many more. These tests are simple and less representative of the supercomputer, hence its position in \autoref{fig:pyramid_evaluation}.
    
    \item Kernel Benchmarks --- compute specific dedicated intensive calculations of actual scientific applications are executed to test the coupling and synergy between various components. The kernel benchmarks are supposed to be simple yet represent a small fragment of scientific applications. For example, SPEC~\cite{juckeland2014spec} (tests shared-memory parallelization scheme), BTIO~\cite{btio} (MPI operations with I/O operations), high-performance Linpack (HPL, computes basic linear algebra operations)~\cite{HPL}, high-performance conjugate gradients (HPCG, performs the conjugate gradients algorithm)~\cite{dongarra2016high} and more. The kernel benchmarks are slightly complex and more representative, as seen in \autoref{fig:pyramid_evaluation}.
    
    \item Mini/proxy apps --- The kernel benchmarks fail to unveil the supercomputer's capabilities in real scientific applications~\cite{hplisbad}. For example, contrary to the HPL benchmark (a highly optimized compute-bound kernel), real scientific applications are usually multi-bound and not necessarily as optimized or as highly scalable as these benchmarks. This discrepancy only increases in time due to the rapid advancement in computation power and a relatively slow advancement in memory bandwidth and latency. As a result, proxy or mini-apps were devised over the years to bridge the gap between kernel benchmarks and the necessary evaluation of a supercomputer~\cite{scalingthesummit, coralbenchmark, bland2012titan}. These mini-apps attempt to mimic a portion of the computational workflow of scientific applications. For example, AMG2013~\cite{amg} mimics a multi-grid linear system solver for unstructured mesh in physical applications; QuickSilver~\cite{richards2017quicksilver} mimics a portion of the Mercury code~\cite{mercury}; and \LULESH{}~\cite{luleshwebpage} that attempts to mimic a portion of ALE3D~\cite{noble2017ale3d}.
\end{enumerate}

Nonetheless, most mini-apps still contain intrinsic differences compared to scientific applications. For example, mini-apps do not necessarily couple multiple scientific domains, which is common in scientific applications, especially in their corresponding physical simulations known as multi-physical simulations. Coupling multiple domains under one application is crucial as it increases memory consumption --- as more data structures are needed to describe the problem. Hence, the required amount of synchronization (MPI) operations that depend on the data structures per cycle increases. Furthermore, in some cases, coupling several domains might heavily change the scheme of the code. As a result, the scaling capabilities of mini-apps hardly reflect their counterpart application.

\subsection{\LULESH{}}
The previously mentioned, well-known \LULESH{} mini-app is highly simplified and hard-coded to simulate the Sedov-Taylor problem~\cite{sedovtaylor} --- it solves hydrodynamic equations (Navier-Stokes) explicitly via an unstructured grid. Hence, it is latency and bandwidth bound~\cite{wen2018profdp}. In addition, \LULESH{} supports multiple parallelization schemes and programming paradigms: MPI, OpenMP, CUDA, OpenACC and OpenCL~\cite{luleshwebpage}. \LULESH{} was developed as part of one of the five challenge problems defined by the DARPA UHPC program~\cite{darpa, luleshwebpage} and attempts to mimic a small portion of various multi-physics applications (specifically, ALE3D~\cite{noble2017ale3d}) that consume up to 30\% of the computing resources of the DoD and DoE~\cite{openmpgpululesh, luleshtuningale3d, hornung2011hydrodynamics}. 

Due to its popularity, \LULESH{} is commonly used for testing new hardware, applying optimizations, parallelization schemes, APIs, and more~\cite{scalingthesummit, openmpgpululesh}. For example, Laney et al.~\cite{laney2013assessing} assessed the effects of data compression on the performance of \LULESH{} and several other benchmarks --- as they are the most representative of real scientific applications. Furthermore, Bercea et al.~\cite{openmpgpululesh} evaluated the effectiveness of the OpenMP version on a GPU.

Nevertheless, there are still many crucial distinctions between \LULESH{} and other common scientific applications such as the ALE3D (which \LULESH{} is based upon). For example, contrary to \LULESH, ALE3D simulates multi-physical phenomena. Therefore, ALE3D has a higher memory consumption corresponding to the number of physical quantities necessary to describe multi-physical phenomena. Moreover, multi-physical simulations also contain additional calculations for each cycle (corresponding to the additional equations), such as multi-phase transitions, thermal, advection, chemistry-related calculations, implicit schemes, and many more. These disparities were expressed and even quantified in~\cite{luleshtuningale3d}. \LULESH{} was tuned and optimized on modern hardware, which resulted in an average of 60\% decrease in total execution time compared to an unoptimized run. However, applying the same optimizations to ALE3D results in only a 20\% time decrease for the same problem, suggesting that although \LULESH{} provides a reasonable benchmark for multi-physics applications, it is far from representing the actual nature of such applications in supercomputers. 

Therefore, similarly to kernel benchmarks, most mini-apps also fail to unveil the true scaling capabilities of supercomputers regarding common scientific applications such as the ALE3D, which, as mentioned, represents 30\% of calculations in the DoD and DoE.

\subsection{Contribution --- \backus{}} \label{sec:backusintro}

We propose a scalable mini-app benchmark framework, namely \backus{}, that will attempt to bridge the gap between benchmarks and actual scientific applications --- or at the very least, provide a framework that will allow bridging this gap. The essence of this benchmark is to create a framework or infrastructure that can be easily expanded to include additional numerical algorithms, physical models, and alternative parallelization schemes with a special focus on scalability, i.e., the ability of this benchmark to maintain a scalable speedup and short execution times, especially compared to currently available benchmarks.

\backus{} utilizes Modern Fortran~\cite{metcalf2004fortran} as the programming language. Modern Fortran provides excellent, if not the best, execution times for numerical calculations compared to most programming languages while also enabling modern object-oriented design patterns~\cite{CFortrancompare1, isfortranrelevant,gustafson1988reevaluating}. Although object-oriented design patterns provide valuable concepts that help the code be more maintainable, easy to modify, and expand (which is beneficial for \backus{} as a framework), it does not go hand to hand with performance. Thus, \backus{} employs a performance object-oriented programming paradigm that aims to utilize design patterns without harming the performance, i.e., favoring performance in the trade-off between design patterns and high-performance. Moreover, there are many legacy codes (that are still widely used) written in Fortran~\cite{loh2010ideal, fortranlegacy,fortranisimportant} (hence its popularity among scientists), contrary to the number of benchmarks in Fortran.  %Please see \ref{chap:foundations} for more details.

\backus{}'s infrastructure is composed of two independent layers that complement one another. The first and most important layer is the \textit{kernel}. This layer implements the basic data structures, for example, a 3D array that can be accessed via a function. Moreover, this layer provides a transparent interface to the distributed parallelization scheme via transparent parallel objects implemented with the MPI-3 standard that further supports \backus{} scaling capabilities. The layer is independent of the other layer and can be applied in different codes and projects. The second layer, the \textit{application}, contains physical quantities and other related calculations, such as equation-of-state (EoS), and provides an abstraction layer to the data structures implemented in the \textit{kernel} layer. Furthermore, the \textit{application} utilizes the physical quantities to implement the physical and numerical models. The \textit{application} layer is generic and can be replaced or added by any physical model, relying on the \textit{kernel} as an infrastructure.

Currently, \backus{} implements 1D, 2D and 3D multi-material hydrodynamic scheme. The hydrodynamic scheme is based on the well-known simplified arbitrary Lagrangian Euler (SALE) code~\cite{SALE, SALE3d}. The SALE's simple structure is easy to follow, maintain, modify and expand. In order to support \backus{} as a framework, the numerical calculations and functions in SALE were adapted to object-oriented design patterns and optimized to modern architectures and compilers. 

\begin{figure}

  \centering
  %%%%%%%%%%%%%%%%%%%%%%%%%%%%%% STRONG TIME
\begin{subfigure}[b]{0.24\textwidth}

%%%%%%%%%%%%%%%%%%%%%%%%%%%%%% STRONG TIME
    \includegraphics[width=\textwidth]{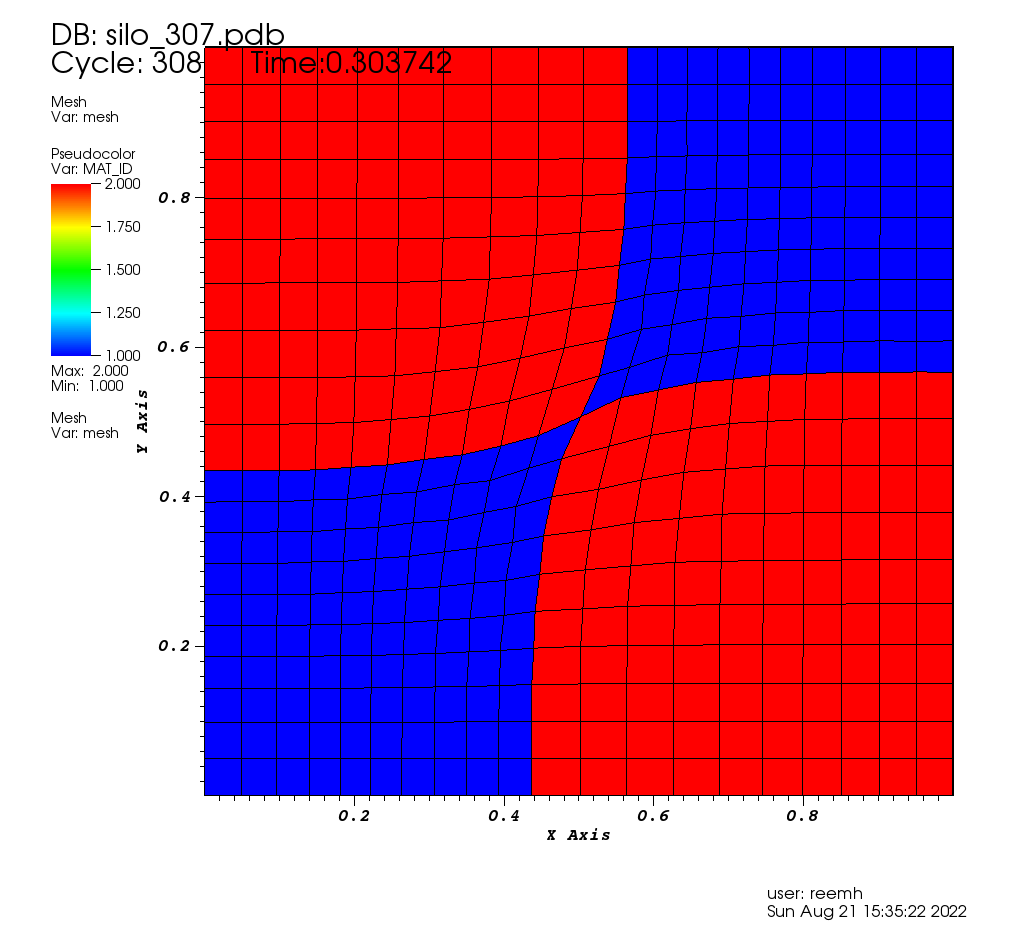}
    
  \subcaption{2D Lagrangian Sod simulation.}
  \label{fig:sod_lagrange}
\end{subfigure}
\hfill
% STRONG EFFICIENCY
\begin{subfigure}[b]{0.24\textwidth}
    \includegraphics[width=\textwidth]{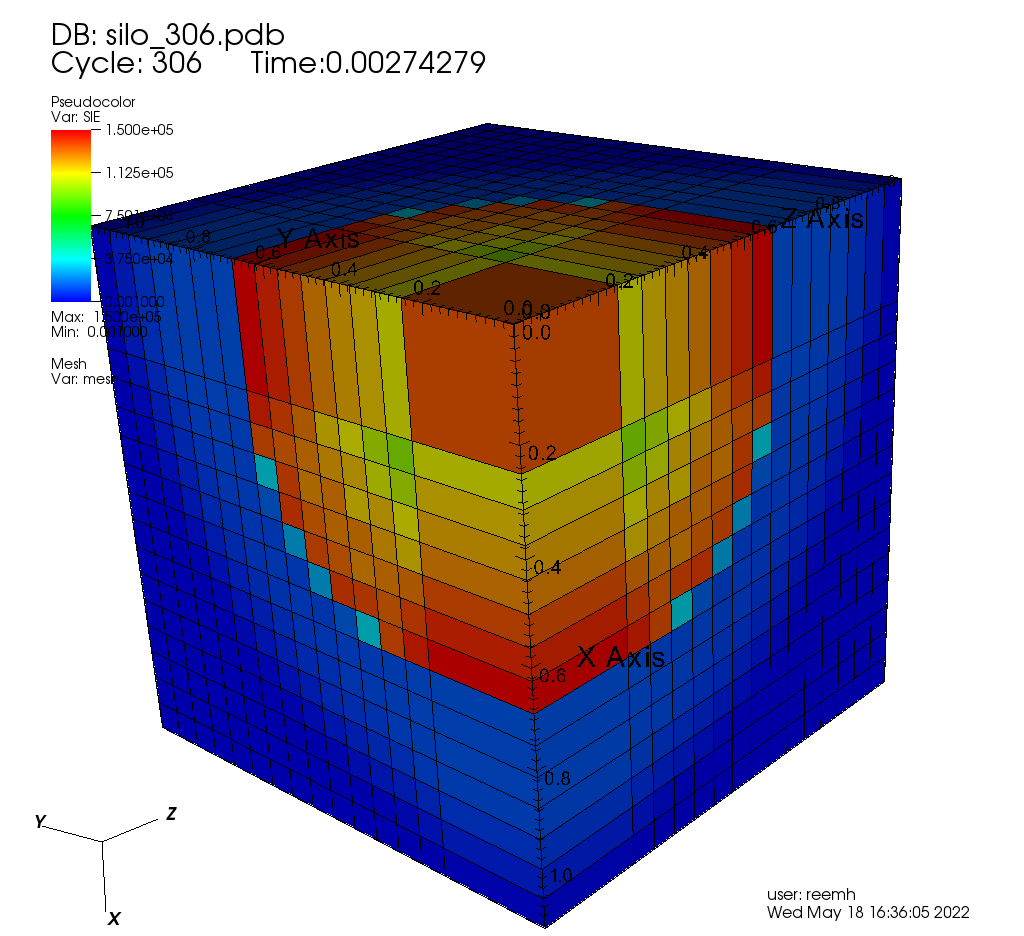}
    \subcaption{3D Eulerian Sedov-Taylor simulation.}
    \label{fig:sedov_euler}
    
\end{subfigure}
\caption{\backus{} simulation of Sod and Sedov-Taylor problems.}
    \vspace{-3ex}

\end{figure}

In order to test the scaling capabilities, it was compared to the well-known hydrodynamic mini-app --- \LULESH{}. As mentioned, \LULESH{} is extensively used, especially for evaluating supercomputers, which makes it highly suitable for evaluating \backus{}. Moreover, both codes perform similar calculations: both solve the Navier-Stokes equations in 3D; can simulate the Sedov-Taylor problem (presented in~\autoref{fig:sedov_euler}); apply the ideal gas equation-of-state and more. % A solution of Sedov-Taylor problem in Eulerian rezoning can be seen in~\autoref{fig:sedov_euler}.

Nevertheless, \backus{} is compared to \LULESH{} only in terms of scaling capabilities, as the main goal of \backus{} is to provide a scalable framework that can be expanded easily, especially for next-generation supercomputers in the exascale era. Contrary to \LULESH{}, \backus{} can also solve additional physical problems such as Sod shock-tube~\cite{sod1978survey} (as seen in~\autoref{fig:sod_lagrange} for Lagrangian rezoning), spherical divergent shocks (Noh)~\cite{noh1983artificial} and more. In addition, \backus{} can simulate problems in 1D, 2D, and 3D, with more than one material (hence, multi-material) and its corresponding equation-of-state. Furthermore, \backus{} solves the equations via a structured grid, with Lagrangian or Eulerian (with an advection phase) rezoning. Thus, providing a wider range of problems that it can solve. The code-features of \LULESH{} and \backus{} are compared and presented in \autoref{table:luleshbackuscomparison}.

% \autoref{table:luleshbackuscomparison} summarizes the comparison between \backus{} and \LULESH{} in different code features.

\begin{table}
\centering
    \begin{tabular}{l c c r}
    \toprule
    \textbf{Feature} &  \textbf{\LULESH{}} & + / & \textbf{\backus{}}  \\
    \midrule
    Mesh & Unstructured & /  & Structured
    \\[3pt]
    Interoperability & -- & / & C, C++
    \\[3pt]
    Parallelization & Blocking & + & Non-blocking
    \\[3pt]
    Paradigm & Variable-oriented & / & OOP
    \\[3pt]
    Code Structure & \begin{tabular}{@{}c@{}}Simplified \\ hard-coded\end{tabular} & / & SALE
    \\[3pt]
    \begin{tabular}{@{}l@{}}Numerical \\ kernels\end{tabular} &  Implicit & + & \begin{tabular}{@{}r@{}}Lagrangian rezoning \\ Eulerian rezoning \\ Advection\end{tabular}
    \\[4pt]
    Equations & Hydrodynamics & + &  \begin{tabular}{@{}r@{}}Multi-material \\ Multiple EoS\end{tabular}
    \\[4pt]
    Dimension & 3D & + & 1D and 2D
    \\[4pt]
    Problems & Sedov-Taylor & + & Shock-tube
    \\[3pt]
    \bottomrule
    \end{tabular}
    \\[3pt]
    \caption{Comparison between \backus{} and \LULESH{} code features. The "+" refers to additional features, and "/" refers to different features.}
    \label{table:luleshbackuscomparison}
    \vspace{-3ex}
\end{table}

\section{Foundations of \backus{}} \label{sec:foundations}

A modern, scalable scientific benchmark, as a framework, should be readable, maintainable, extensible, well-tested, well-documented, portable to modern architectures, and most importantly --- be constructed in a way that will allow it to take advantage of the optimal parallelism schemes~\cite{feathers2004working}. A benchmark that copes with these requirements must, on the one hand, be written in a programming language with high-enough abstraction capabilities while, on the other hand, be as light-weighted as possible. This section presents the foundations of \backus{}, that enable such a scalable benchmark framework, based on~\cite{rusanovsky2019backus}.

\subsection{High-performance Oriented Design and Code} \label{sec:modern_fortran}
  
  The programming languages' abstraction--performance trade-off consists of very fast languages in one extreme and high abstraction level ones in the other. An example of the former languages is Assembly, which is very fast since it is close to the machine code and does not require many translations. However, programming in Assembly requires a very high effort and expertise and does not support OOP. Most importantly, it is not suitable as a benchmark since it does not represent most applications written in higher abstraction level programming languages. An example of the latter language is Python, which is very expressive and easy to use and maintain. Furthermore, it supports OOP but is much slower since it translates the programming abstractions into machine code in run-time.
  Between these two extremes, there is a range of languages, starting from Assembly, with a decreasing performance power coupled with increasing expressive power. This range includes, among others, Assembly, Fortran 66, Fortran 77, Fortran 90, C, Modern Fortran, C++, and Python.
  Fortran is used in many benchmarks and codes~\cite{fortranbenchmark, fortrancode, computationalphysicsfortran, fortranlegacy,fortranisimportant} mainly due to its low execution times while supporting sufficient programming abstraction.
  
  Fortran underwent a major revision in 2003 and subsequently was re-branded as Modern Fortran. Modern Fortran supports object-oriented design patterns, thus enabling system design through modern design concepts \cite{hanson2013numerical, gorelik2004object}, while still benefiting from the fast execution time of the language. In addition, the language features the basic concepts of object-oriented programming (OOP): objects, inheritance, polymorphism, dynamic type allocation, type-bound procedures, and many more. 
  In 2008, Modern Fortran was updated to support additional sub-modules, coarrays, \emph{DO CONCURRENT} construct that hints to the compiler that a loop can be parallelized, and more. The latest version of Modern Fortran was released in 2018 and included further interoperability with C and additional parallelization features in Fortran.
  
  Over the years, Fortran and C++ were compared to decide which language is best suited for large-scale codes, both in execution and in numerical capabilities, e.g., in multi-physics and scientific applications~\cite{isfortranrelevant,moreira1998comparison, CFortrancompare1, CFortrancompare2, CFortrancompare3, CFortrancompare4}. 
  The initial conclusion was that even with the faster execution times of Fortran, C++ is preferred due to the poor support in OOP design patterns and the programming effort Fortran requires.
  Then, Modern Fortran was upgraded to support more design patterns, resulting in a preference for Modern Fortran over C++. Currently, Fortran is regarded as the indisputable leader for carrying out highly optimized scalable scientific applications in the shortest execution times --- even more so since the introduction of Modern Fortran.
  Moreover, Fortran nowadays provides excellent interoperability with C and C++. Thus, applications written in Fortran, specifically with Modern Fortran object-oriented design patterns, can include modules written in C and C++ --- providing the best from both worlds.
  
 Although Modern Fortran does not fully support all object-oriented features such as generic types, pointers, and polymorphic arrays, these design patterns are not necessarily needed. Instead, they may be replaced with other available programming solutions --- which usually require slightly higher programming effort. On the other hand, the lack of available abstractions offered in Modern Fortran guarantees the relatively thin implementation for OOP support. This abstraction deficiency demands higher programming effort but allows for an execution time reduction in general and in scientific codes specifically~\cite{arabas2014formula}. % More details on OOP conflicts in high-performance scientific applications and benchmarks and their corresponding workarounds in \backus{} are presented in \autoref{sec:OOP_conflicts}.
 
 % OOP is a powerful paradigm, but as a rule of thumb, it does not necessarily conform with performance. For example, abstract classes are great for defining functionality without a concrete implementation (such as EoS calculations). However, if a class that implements an abstract class is called multiple times in a single loop iteration, the memory and computation overhead, derived from the inheritance, becomes significant, which in turn harms the total execution time. In contrast, a simple switch-case clause is thin computationally but lacks the benefits of OOP. Thus, creating a scalable scientific code based on OOP principles must be done carefully and with the right balance between design patterns and scalable execution time.

% \backus{} follows Object-Oriented Parallel Programming (OOPP) ideas which resembles other OOP-based scientific applications such as \cite{pinho2014object,calvin2002object}, in which objects are intrinsically parallel. Nevertheless, developing an efficient and scalable scientific code may and usually will violate important OOP principles, such as object encapsulation, functional independence, and modularity. This trade-off is investigated with caution, but an OOP solution is often preferred if it does not come with substantial running-time growth. Other works show OOPP variants that emphasize the importance of introducing OOP into HPC applications \cite{fan2017towards,radenski1998object}, and thus keeping the benefits of OOP (without breaking too-many OOP principles) while preserving the efficiency required from modern scientific codes.

\subsection{Transparent Parallel Class} \label{sec:oopp}

Since the introduction of object-oriented programming, principles such as modularity, abstraction, encapsulation, and hierarchy have been encouraged. These principles make it easier to maintain and modify large software systems.
Modular programming requires that the code follow decomposability and composability, among other concepts, which promote coupling data structures with their related functions within classes. These concepts allow new scientific applications' developers and users to easily understand these classes' functionality and the code's general structure, subsequently expanding them with additional functionalities~\cite{stroustrup2000c++, rouson2011scientific}.

Modularity is beneficial for developing and composing new scientific applications and for decomposing conceptually independent implementations from each other. This is significantly useful for separating the implementations of parallel modules from the rest of the scientific application. Following this idea, the parallel implementation is abstracted by a class that can be updated locally instead of along the entire code to support, e.g., a new MPI standard.
Abstraction of this sort also allows to optimize the parallel scheme and tune it tightly to given hardware in a single code region.

OOP is a powerful paradigm, but as a rule of thumb, it does not necessarily conform with performance. For example, abstract classes are great for defining functionality without a concrete implementation (such as EoS calculations). However, if a class that implements an abstract class is called multiple times in a single loop iteration, the memory and computation overhead, derived from the inheritance, becomes significant, which harms the total execution time. In contrast, a simple switch-case clause is thin computationally but lacks the benefits of OOP. Thus, creating a scalable scientific code based on OOP principles must be done carefully and with the right balance between design patterns and scalable execution time.

\backus{} follows Object-Oriented Parallel Programming (OOPP) ideas which resemble other OOP-based scientific applications such as \cite{pinho2014object,calvin2002object}, in which objects, specifically the data structures, are intrinsically parallel. Thus, developers can synchronize the data structures with ease. Nevertheless, developing an efficient and scalable scientific code may and usually will violate important OOP principles, such as object encapsulation, functional independence, and modularity. This trade-off is investigated with caution, but an OOP solution is often preferred if it does not come with substantial running-time growth. Other works show OOPP variants that emphasize the importance of introducing OOP into HPC applications \cite{fan2017towards,radenski1998object}, and thus keeping the benefits of OOP (without breaking too-many OOP principles) while preserving the efficiency required from modern scientific codes.

\section{Implementation}
% \rh{MANY CITES}
\backus{} is designed as a computational infrastructure for scalable multi-material and multi-physical numerical benchmarks. \backus{} is written in Modern Fortran~\cite{metcalf2011modern} (See \autoref{sec:modern_fortran}) and implemented using simple yet practical design patterns from OOPP (See \autoref{sec:oopp}). Further implementation details and important concepts are presented.

\subsection{High-performance Code Written in OOP} \label{sec:OOP_conflicts}
As discussed in \autoref{sec:foundations}, the foundations of \backus{} paradigm is a balance between high programming effort and fast codes to easier to use and slower codes. \backus{} as a benchmark was developed following general OOP principles, with exceptions that favor high-performance over OOP paradigms. Following are representative exceptions for each OOP main principle.

\begin{enumerate}[itemindent=-4pt,topsep=0pt,itemsep=-1ex,partopsep=0ex,parsep=1ex]
    \item Encapsulation: Usually, only the object has the authority to access and modify its data attributes directly. The access to the object's data from other objects is done via public methods such as get/set. These functions might incur significant time and memory overheads since they require copying and setting the entire data attribute for each access. Since data attributes in scientific applications may be very large, this functionality is not desired. A workaround adapted in \backus{} is to use \textit{Point\_to\_data} functions, which return a pointer directly to the data attribute from outside the object. This eliminates the need to copy the data and serves as a \say{hand-shake} between the object that owns the data attribute with the code segment that accesses the data.
    
    \item Polymorphism: Modern Fortran does not support array of polymorphic types. This functionality is useful for defining the boundary conditions scheme for each edge of the mesh. A fully polymorphic solution would have had an abstract base class of cell boundary conditions and concrete implementations of cell boundary conditions' calculations such as slip cell and free surface classes. Since cell boundary conditions on a given edge of the mesh do not change during the simulation, it might be convenient to initialize the specific cell boundary condition class at the beginning of the program and store an array with a polymorphic pointer to the specific cell boundary condition class for each edge of the mesh. A simple workaround would be to define a wrapper type with a single attribute, a polymorphic pointer, to the base cell boundary conditions class. An array of these wrapper types can implement the edge boundary conditions array. 
    
    \item Abstraction: Following this principle, objects should be exclusively responsible for their functionalities and data attributes. For example, it would be preferable for each physical quantity class in \backus{} to contain a method called calculate. However, since each physical quantity's calculation should accept different quantities as parameters, and since the calculation depends on the actual physical/numerical scheme, these calculations are kept in a more global place. For example, \textit{Calculate\_density} is placed in \textit{Hydro} and not within the density class.
    
    \item Inheritance: This principle allows classes to inherit from base classes and reuse or overload some basic functionalities from that base class. This property reduces programming time and requires more thorough computation and data analysis. However, since overloading functions come with an overhead cost, this functionality should be used cautiously. In \backus{} there are scenarios in which inheritance is used and functions overloading is carried out, and in other cases in which a switch--case clause is used to determine the desired functionality (e.g., for determining whether to perform 2D or 3D rezoning).
\end{enumerate}

\subsection{\backus{} as an Infrastructure}
\begin{figure}
\includegraphics[trim={2.8cm 7.8cm 1.8cm 4cm},clip,width=9cm]{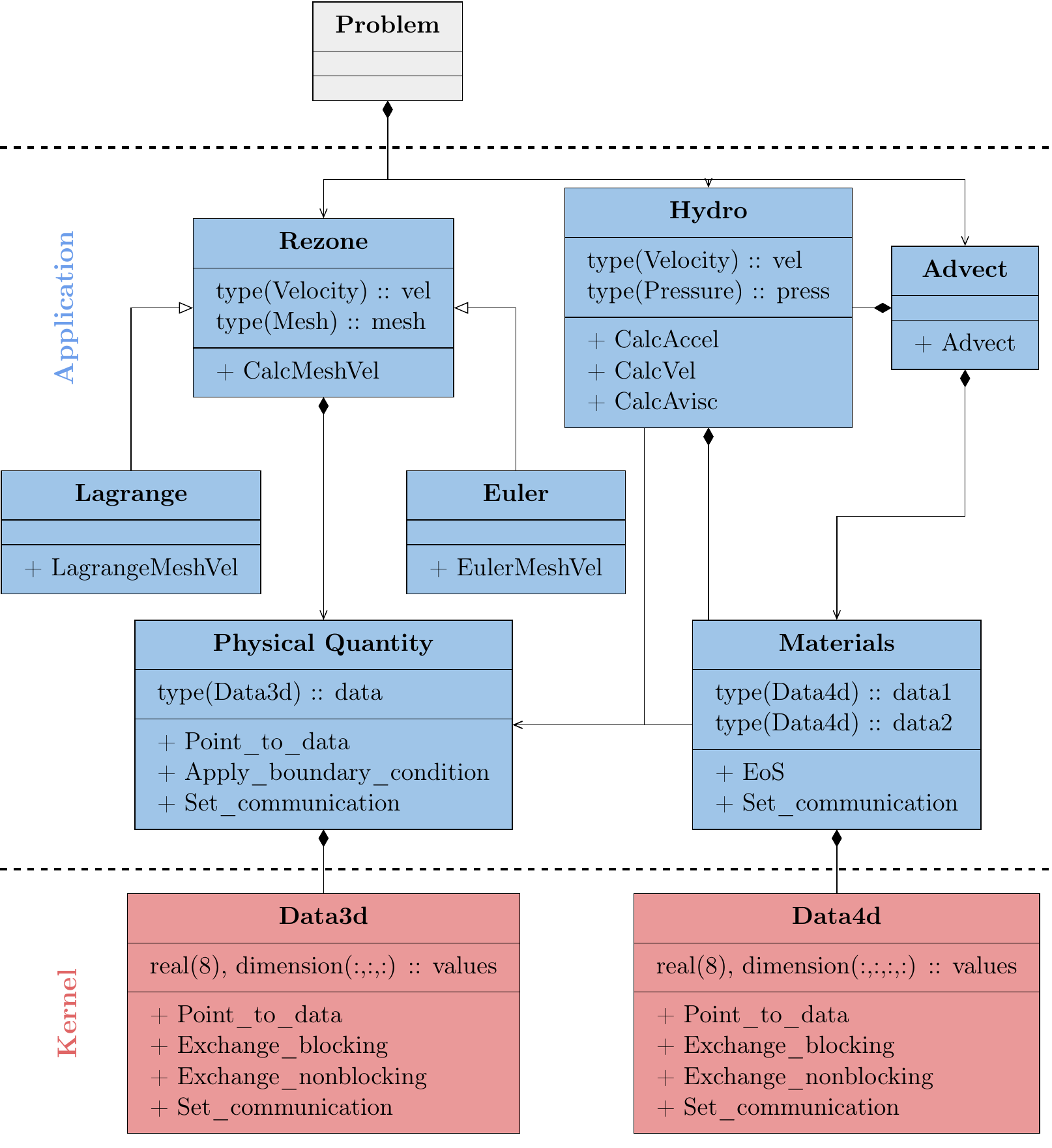}
\caption{\backus{} schematic UML. The classes in blue are associated with the \textit{application} layer, while classes in red are associated with \textit{kernel}.}
\label{fig:uml}
    \vspace{-3ex}

\end{figure}

\backus{} is implemented as two layers that are independent yet complementary to one another: the \textit{kernel} layer and the \textit{application} layer. This design provides a simple approach for expanding and modifying each of the two layers independently.
An overview of the UML of \backus{} is presented in \autoref{fig:uml}.

\subsubsection{The Kernel}
Handles the computational building blocks for the structured grid calculations. It consists of the data structures that are encapsulated in the classes \emph{Data3d} and \emph{Data4d} (as presented in \autoref{fig:uml}), and the parallelization schemes that are implemented in the \textit{Communication} and \textit{CommParams} classes (as presented in \autoref{fig:parallel_object}, and will be explained in the following section). The interface between the data structures and the parallelization schemes is given by functions such as \emph{Point\_to\_data} and \emph{Exchange\_blocking}. Furthermore, the \textit{kernel} layer contains third-party libraries, enums, linked list implementation, and more.
%The purpose of this class is to allocate the arrays or other data structures and execute the distributed parallelization scheme -- which will be explained in the next section. In addition, the class provides a subroutine \emph{Point\_to\_data} that returns a pointer to a 3D or 4D array. The subroutine further helps encapsulate the data structure and allows future modifications to the data structure -- ensuring that the application will always receive an array while implementing the data structure differently. 

\subsubsection{The Application}
Handles all the physical-related computations such as thermodynamics, hydrodynamics, mesh rezoning, advection, and future models (as seen in \autoref{fig:uml}). Generally, each physical module should be associated with this layer and be implemented as an independent class. Associating physical modules with classes allows easily expanding these physical modules or overloading them with different schemes. Furthermore, every physical module utilizes different physical quantities represented by pointers to the relevant objects of physical quantities. The physical quantities and their related calculations further assist the encapsulation of the \textit{kernel} layer from the physical models in the \textit{application} layer, enhancing the simplicity and readability of both layers. Moreover, the numerical computations are written with minimal OOP intervention for minimal execution time reduction.

% \begin{figure}
% \includegraphics[trim={2.5cm 10cm 0cm 3.77cm},clip,width=9cm]{Stress_UML.pdf}
% \caption{The UML of \backus{} after the addition of stress model}
% \label{fig:stress_object}
% \end{figure}

% MPI

\subsection{Communication Implementation}\label{sec:mpi3}
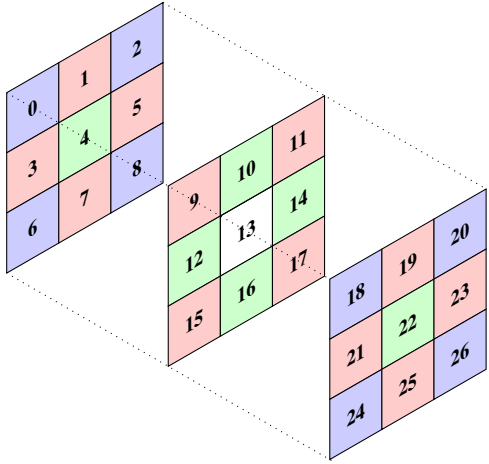
\begin{figure}
    \centering
        \begin{tikzpicture}
 [x={(0.866cm,0.5cm)}, y={(-0.866cm,0.5cm)}, z={(0cm,1cm)}, scale=0.8,
 bluenode/.style={shape=rectangle, draw=blue!20, line width=2,fill=blue!20},
  rednode/.style={shape=rectangle, draw=red!20, line width=2, fill=red!20},
  greennode/.style={shape=rectangle, draw=green!20, line width=2, fill=green!20}
]

    \begin{scope}[canvas is xz plane at y=-3,transform shape]
      %\draw[blue] (0,0) -- (10,0)--(10,10)--(0,10)--cycle;
      \foreach \ii [count = \xi] in {0,1,2}{
            \foreach \jj  [count = \yi]in {0,1,2}{
            \pgfmathsetmacro{\nn}{int(\xi+3*\yi-3-1)}
            
            \ifthenelse{\nn=1 \OR \nn=3 \OR \nn=5 \OR \nn=7}{\node[draw,minimum size=1cm, fill=red!20] (n\nn-3) at (\ii,-\jj) {\textbf{\nn}};}{\ifthenelse{\nn=0 \OR \nn=2 \OR \nn=6 \OR \nn=8}{\node[draw,minimum size=1cm,fill=blue!20] (n\nn-3) at (\ii,-\jj) {\textbf{\nn}};}{\node[draw,minimum size=1cm,fill=green!20] (n\nn-3) at (\ii,-\jj) {\textbf{\nn}};}}
      %\node[draw,minimum size=1cm] (n\nn-1) at (\ii,-\jj) {\nn};
      }
      }

      \end{scope}

        \begin{scope}[canvas is xz plane at y=-6.1,transform shape]
      \foreach \ii [count = \xi] in {0,1,2}{
            \foreach \jj  [count = \yi]in {0,1,2}{
            \pgfmathsetmacro{\nn}{int(\xi+3*\yi-3-1+9)}
            
            \ifthenelse{\nn=9 \OR \nn=11 \OR \nn=15 \OR \nn=17}{\node[draw,minimum size=1cm, fill=red!20] (n\nn-3) at (\ii,-\jj) {\textbf{\nn}};}{\ifthenelse{\nn=13}{\node[draw,minimum size=1cm] (n\nn-3) at (\ii,-\jj) {\textbf{\nn}};}{\node[draw,minimum size=1cm,fill=green!20] (n\nn-3) at (\ii,-\jj) {\textbf{\nn}};}}
    %   \node[draw,minimum size=1cm] (n\nn-2) at (\ii,-\jj) {\nn};
      }
      }
      \end{scope} 

      \begin{scope}[canvas is xz plane at y=-9.2,transform shape]
      \foreach \ii [count = \xi] in {0,1,2}{
            \foreach \jj  [count = \yi]in {0,1,2}{
            \pgfmathsetmacro{\nn}{int(\xi+3*\yi-3-1+18)}
            \ifthenelse{\nn=23 \OR \nn=19 \OR \nn=21 \OR \nn=25}{\node[draw,minimum size=1cm, fill=red!20] (n\nn-3) at (\ii,-\jj) {\textbf{\nn}};}{\ifthenelse{\nn=18 \OR \nn=20 \OR \nn=24 \OR \nn=26}{\node[draw,minimum size=1cm,fill=blue!20] (n\nn-3) at (\ii,-\jj) {\textbf{\nn}};}{\node[draw,minimum size=1cm,fill=green!20] (n\nn-3) at (\ii,-\jj) {\textbf{\nn}};}}
      %\node[draw,minimum size=1cm] (n\nn-3) at (\ii,-\jj) {\nn};
      }
      }
      \end{scope} 
      \draw [dotted] (0,-2.5) -- (0,-8.7);
      \draw [dotted] (-3,-5.6) -- (-3,-11.8);
      \draw [dotted] (3,-2.5) -- (3,-8.7);
    %   \matrix [draw,below left] at ($(current bounding box.south)+(1.1,-2.4)$) {
    %     \node [greennode, label=right:High communication load] {};\\
    %     \node [rednode,label=right:Medium communication load] {}; \\
    %     \node [bluenode,label=right:Low communication load] {}; \\
    %     };

    %   \draw[fill=red!50,opacity=0.3] (n10-1.north east) -- (n5-2.north east) --(n10-3.north east)
    %   --(n10-3.north west)-- (n5-2.north west) --  (n10-1.north west)  ;
    %   \draw[fill=red!50,opacity=0.3] (n10-1.south east) -- (n5-2.south east) --(n10-3.south east)
    %   --(n10-3.south west)-- (n5-2.south west)-- (n10-1.south west)  ;    
    %   \draw[fill=red!50,opacity=0.3] (n10-1.north east) -- (n5-2.north east) --(n10-3.north east)
    %   --(n10-3.south east)-- (n5-2.south east)-- (n10-1.south east)  ;

    %   \draw[fill=red!50,opacity=0.3] (n10-3.north west)-- (n5-2.north west) --  (n10-1.north west) -- (n10-1.south west)-- (n5-2.south west)   --(n10-3.south west) ;        

  \end{tikzpicture}
    \caption{3D communication topology structure of an MPI rank --- 27 neighbors. The 2D communication topology can be derived straightforwardly by limiting the ranks to a 2D rectangle --- ranks 9 until 17, for example.}
    \label{fig:MPI_topology}
    \vspace{-3ex}
\end{figure}

The parallelization scheme in \backus{} distributes the mesh into cubes or tiles and uses ghost cells~\cite{kjolstad2010ghost} for stencil calculations (in which each cell depends on its nearest neighbors). Each MPI process has 27, 8, or 2 neighbors in 3D, 2D, and 1D, respectively. When the processes synchronize in a 3D setting, $O(N^2)$ while $N$ is the problem size, values are sent to the closest face neighbors (presented in green in \autoref{fig:MPI_topology}); $O(N)$ values are sent to the edge neighbors (presented in red in \autoref{fig:MPI_topology}) and $O(1)$ values are sent to the corner neighbors (presented in blue in \autoref{fig:MPI_topology}).

The communication-related operations utilize the MPI-3 standard. The MPI-3 standard introduced important features such as an extension of collective operations to non-blocking versions, wider support for Fortran, C and C++ binding, point-to-point communications, user-defined communication topologies, and thread-safe operations. Each feature contributes to the scalability of a parallel (even hybrid parallelization schemes) framework written in Modern Fortran.

\backus{} intent is to provide a scalable benchmark framework. Therefore, the synchronization overhead should be minimized as much as possible. To comply with this idea, non-blocking communication, and a communication topology were introduced to \backus{}:

\subsubsection{Non-blocking Communication}
Na\"ively, ghost cells should be synchronized across the ranks after each update of a physical quantity. The update process consists of sending and receiving MPI messages to the corresponding neighbors. The communication overhead increases as the number of ranks increases due to the higher probability of a delayed message in the system. 
A possible approach to diminish such a case is by introducing non-blocking communication to the code --- validating that the received messages indeed arrived right before they are being used. The non-blocking communication grants the ability to perform other calculations between the two points of synchronization (send and receive). Thus, if an MPI message is delayed, the program performs other calculations instead of being idle, which minimizes the communication overhead. 
\backus{} implements non-blocking communications via the asynchronous collective MPI operation \emph{Ineighbor\_alltoall}.

\subsubsection{Topology}
In supercomputers, cores on the same node transfer messages via fast shared-memory communication, while cores from different nodes are connected via intercommunication devices such as Infiniband and Ethernet, which are slower by factors or orders of magnitude, respectively.
Therefore, ranks that communicate frequently (for example, the green ranks in \autoref{fig:MPI_topology}) should be assigned to cores that reside in the same node --- thus, lowering the communication overhead.

In \backus{}, this idea is implemented via the communication topology function that was presented in MPI-3, \emph{MPI\_Dist\_graph\_create\_adjacent}. This function creates a graph topology between the appropriated processes and initializes the communicator. Furthermore, the function receives a weight array that hints at the communication load between the ranks. In turn, the function may migrate the ranks (change their original identifiers) such that the ranks that heavily communicate will reside in the same node.

\begin{figure}
\includegraphics[trim={3.3cm 15.25cm 3.1cm 5.8cm},clip,width=8cm]{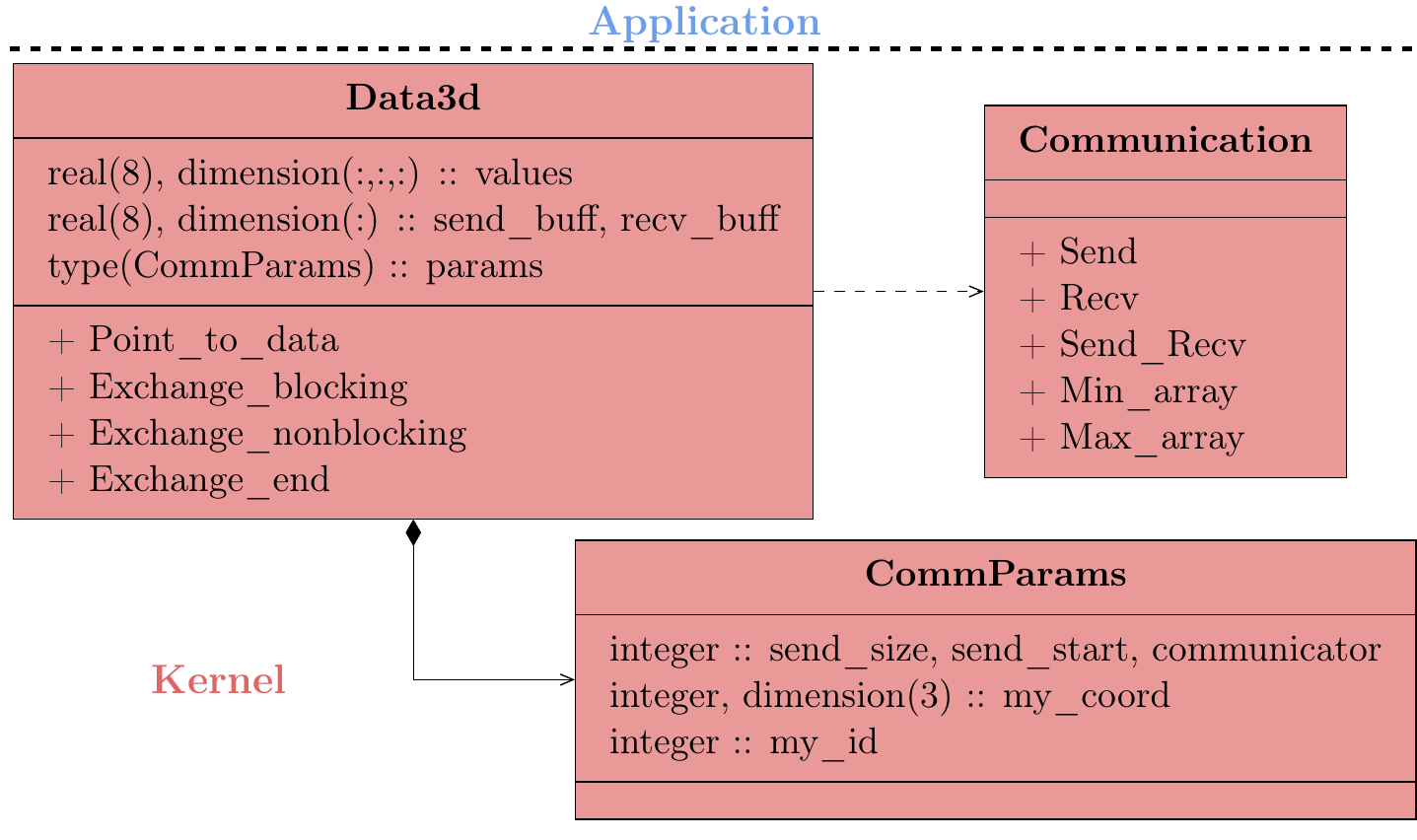}
\caption{\backus{} UML of the parallel object in \backus{}.}
\label{fig:parallel_object}
    \vspace{-2ex}

\end{figure}

The parallelization scheme is associated with the \textit{kernel} layer and is implemented via three classes (see \autoref{fig:parallel_object} a schematic overview of the three classes): 

\begin{enumerate}[itemindent=-4pt,topsep=0pt,itemsep=-1ex,partopsep=0ex,parsep=1ex]
    \item Optimized communication --- \emph{Communication} class: Handles and abstracts the communication operations. Each function implements a communication operation, for example, \emph{Send}, \emph{Recv} and \emph{Send\_Recv}. These functions also handle the synchronization process of the ghost cells, which, as mentioned, is implemented via the \emph{MPI\_Ineighbor\_alltoall} function~\cite{mpialltoall}. 
    
    \item Topology and ranks --- \emph{CommParams} class: Initializes the communicator with the \emph{MPI\_Dist\_graph\_create\_adjacent} function. Furthermore, the class handles the relevant parameters that describe the MPI messages, i.e., size of messages, the communication topology (communicator), and more. %  In \backus{}, physical quantities are differentiated by their position in the computational mesh -- a vertex physical quantity and a cell physical quantity. Hence, the indices of the actual values that should be sent to the appropriate neighbour is different -- vertex quantities indices are located in $(i \mp 2,j \mp 2,k \mp 2)$ and cell quantities in $(i \mp 1,j \mp 1,k \mp 1)$
    
    \item Transparent communication --- \emph{Data3d}/\emph{Data4d} class: Implements the basic data structure of \backus{}. As such, it is also an inherently parallel data class that provides a transparent interface for exchanging and synchronizing data across the relevant ranks. Hence, encapsulating the parallel implementation from the rest of the program. The \emph{Data3d} and \emph{Data4d} classes provides three exchanging operations: a blocking operation (\emph{Exchange\_blocking}), a non-blocking send operation (\emph{Exchange\_nonblocking)} and a non-blocking receive operation (\emph{Exchange\_end}).
    In addition, it contains the values of the data and the communication parameters via \emph{CommParams} as a variable. The \textit{CommParams} variable is provided by the physical quantity, thus, hiding the physical quantity type from \emph{Data3d} and \emph{Data4d} classes.
\end{enumerate}

\subsection{A Case Study --- Stress Module}

\begin{figure}
\centering
\begin{subfigure}{0.42\textwidth}
\begin{minted}[fontsize=\notsotiny, linenos, frame=single]{fortran}
type, extends(cell_quantity_t) :: stress_tensor_t
   ...
end type stress_tensor_t
\end{minted}
\caption{The class describing the new stress tensor quantity, it inherent from the cell quantity base type.}
\label{fig:stress_qunatity}
\end{subfigure}
\vspace{0.1cm}

\begin{subfigure}{0.42\textwidth}
\begin{minted}[fontsize=\notsotiny, linenos, frame=single, firstnumber=4]{fortran}
module material_module
type, extends(material_base_t) :: material_t
  integer, dimension(:) :: stress_module_type 
  ! mat dependent module type (steinberg, perfect, etc)
  contains
    procedure :: Apply_stress_module 
    !calculations of the stress yield and shear
end type material_t
contains
subroutine Apply_stress_module(this, shear, yield)
  implicit none
  class(material_t), intent(in out) :: this
  type(shear_modulus_t), pointer, intent(in out) :: shear 
  ! the shear modulus for every material
  type(stress_yield_t), pointer, intent(in out) :: yield 
  ! the yield for every material
  integer :: i, j, k, m
  do m=1, nmats ! loop over all materials
  do k=1, nz ! in the loops we calculate shear 
  ! and yield for the steinberg model
    do j=1, ny
      do i=1, nx
        ! yield and shear calculations using the
        ! Steinberg stress hardening model
        eta = density0(i,j,k)/density(i,j,k)
        shear(m,i,j,k) = shear0+GP_steinberg(m)* &
        pressure(i,j,k)*eta**(1d0/3)+GT_steinberg(m)* &
        (temperature(i,j,k)-init_temperature(i,j,k))
        yield(m,i,j,k) = min(Y0(m)*(1+BETA_steinberg(m)* &
        plastic_strain(i,j,k)) ** N_steinberg(m) &
        ,YIELD_max(m))*(shear(m,i,j,k)/shear0(m))
      end do
    end do
  end do
  end do
  !the MPI synchronization 
  call shear%Exchange_virtual_space_blocking() 
  call yield%Exchange_virtual_space_blocking()
end subroutine Apply_stress_module
...
end module material_module
\end{minted}
\caption{This code describes adding a new material-dependent calculation in the material class. The numerical stress model calculations of shear and yield use the Steinberg stress hardening model~\cite{steinberg1989constitutive}.} \label{fig:stress_calculation}
\end{subfigure}
\vspace{0.1cm}

\begin{subfigure}{0.42\textwidth}
\begin{minted}[fontsize=\notsotiny, linenos, frame=single, firstnumber=40]{fortran}
type :: stress_model
  type(shear_modulus_t) , pointer :: shear
  type(stress_yield_t)  , pointer :: yield
  type(stress_tensor_t) , pointer :: stress_tensor
  contains
    procedure :: Calculate_stress
    procedure :: Update_acceleration
end type stress_model
\end{minted}
\caption{The class described the new physical stress module. It contains two main functions, a function that calculates the stress in the current time step and a function that update the acceleration values using the updated stress tensor.}
\label{fig:stress_module}
\end{subfigure}
    \caption{A schematic code of stress physical model addition to \backus{}.}
    \label{fig:stress_code}
    \vspace{-3ex}
\end{figure}

Physical attributes such as stress play an important role in describing the motion of solids in hydrodynamic codes. In non-compressible, visco-elastic, or visco-inelastic fluids, forces between the different layers of the fluid result in stress forces that oppose the fluid flow. The stress is strain and velocity dependent, influencing the fluid's momentum distribution and other hydrodynamic properties~\cite{rivlin1948hydrodynamics}. Hence, the stress-strain-velocity relation, which is material-dependent, is crucial for describing the motion of visco-fluids correctly. 

As a case study, the following workflow describes how the stress module can be easily added to \backus{} relatively. Firstly, adding new physical quantities, such as the stress tensor that inherits from the cell quantity base class, to the \textit{application} layer (as seen in~\autoref{fig:stress_qunatity} in lines 1--3). Afterward, implementing the yield and shear modulus calculations~\cite{steinberg1980constitutive} to the \textit{application} layer. The physical calculations are added as new procedures in the materials class as seen in~\autoref{fig:stress_calculation} from lines 4--44. In lines 40--41, the parallelization synchronization is done using a blocking manner. Finally, adding the physical stress calculation~\cite{wilkins1963calculation}, and the correction to the acceleration due to stress, in a new stress model class in the \textit{application} layer, as seen in the code in~\autoref{fig:stress_module} in lines 40--47.
% \end{enumerate}

\section{Results}

% \rh{why we have mpi, and --- scale outer node communication}

To evaluate the potential of \backus{} as a scalable benchmark, it is compared to \LULESH{}~\cite{karlin2013lulesh} --- the well-known mini-app benchmark. \LULESH{} is a hydrodynamic code, highly optimized, that demonstrates excellent scaling capabilities. Moreover, both codes can simulate the same physical problem --- 3D Sedov-Taylor, providing comparable numerical workloads. Thus, it is the most suitable for comparing and evaluating \backus{}. 

Most (if not all) applications that aim to be executed in supercomputers are parallelized with MPI due to the need to scale enormously beyond a single node. However, the introduction of heterogeneous systems poses a challenge in obtaining maximum performance via MPI exclusively. As a result, the newer standards of MPI (especially MPI-3, the standard that is used in \backus{}) supply comprehensive support and even inherently supports shared-memory environments, hence its popularity among hybrid parallelization paradigms (MPI+X)~\cite{he2002mpi}. Nevertheless, evaluating the distributed parallelization scheme scaling capabilities exclusively (MPI) is a necessary step that must be taken independently to MPI+X. 
Therefore, although \LULESH{} implements several parallelization paradigms, the MPI version was exclusively enabled and compared to \backus{}. At the same time, \backus{} could be easily extended to other paradigms without needing the transparent parallel class (for example, inserting an OpenMP directive in line 21 in \autoref{fig:stress_calculation}). 
%Furthermore, the new standart also inherently suppost such parallisem \cite{}.  
%Combined with the wide use of heterogeneous nodes in supercomputers, MPI provides the ideal setting for testing supercomputers' compatibility with said paradigms, as it is the basic foundation of most, if not all, scientific applications. 

\backus{} can simulate physical problems with Lagrangian or Eulerian rezoning (with an advection phase), contrary to \LULESH{} which can simulate with Lagrangian rezoning exclusively. Thus, both rezoning must be considered to evaluate the full scaling capabilities of \backus{}. Executing \backus{} with Lagrangian rezoning provides a more comparable execution time to \LULESH{} while Eulerian rezoning provides a more stressful and realistic calculation as the advection phase results in additional numerical calculations and synchronization points. 

Another point of reference between the two physical schemes is that in Lagrangian rezoning, there are approximately 45 calls to a function in \emph{Communication} i.e., 45 MPI collective operations per cycle, while Eulerian rezoning performs 60 calls per cycle. Each call to a function performs a communication operation of a single physical quantity. In contrary, \LULESH{} performs approximately 5 calls to a similar function per cycle (basic \emph{MPI\_IRecv} and \emph{MPI\_ISend} functions), while each call transfers multiple physical quantity at once. Moreover, as \backus{} is based on the multi-material SALE scheme, it contains additional material-related quantities. As a result, \backus{} not only performs more communication operations but sends more data per cycle. Nevertheless, due to the utilization of non-blocking collective communication and topology distribution via the MPI-3 standard, \backus{} achieves comparable results to \LULESH{}.

The weak and strong scaling tests are used to compare \backus{} and \LULESH{}. The strong scaling test is performed by creating a fixed size problem, considering it has a sufficient workload and increasing the number of cores. The weak scaling test is performed by creating a fixed size problem per core and increasing the number of cores. Both scaling tests are important to understand the scaling capabilities of the parallelization scheme and, in turn, the communication overhead of the code. The efficiency is calculated as follows:
\[
Efficiency =
\begin{cases}
\frac{Time_1}{Time_n} & \text{Weak scaling,} \\
\frac{Time_1}{n \cdot Time_n} & \text{Strong scaling} 
\end{cases}
\]
whereas $Time_1$ is the execution time of a single core, $Time_n$ is the execution time where $n$ is equal to the number of cores.

In Lagrangian rezoning, the amount of numerical operations per cycle is constant. Thus, the evaluation was carried out over only 20 cycles, without the initialization time (similar to~\cite{lulesh20}). In contrast, in Eulerian rezoning, the number of operations per cycle fluctuates due to the advection phase. Hence, 200 cycles were performed for more precise estimation.

The two codes were compiled with Intel's oneAPI~\cite{oneapi} MPI Fortran (Modern Fortran 2018), and C++ compilers \cite{oneapifortran}. oneAPI provides wide support to multiple programming languages~\cite{oneapifeatures} and parallelization paradigms, especially for heterogeneous nodes, thus allowing further local implementation of MPI+X strategy using Intel's oneAPI OpenMP* \cite{oneapiopenmp}. The codes were compiled with no compiler optimization (O0) to allow a fair comparison of the raw scaling capabilities. The codes were executed on NegevHPC~\cite{negevhpc} on Intel Xeon Gold 6130 compute nodes. Each node has 32 cores and 128GB of RAM.

% whereas 

\subsection{Strong Scale}
\begin{figure}

   \centering
  %%%%%%%%%%%%%%%%%%%%%%%%%%%%%% STRONG TIME
\begin{subfigure}{0.4\textwidth}
\centering
\begin{tikzpicture}

\begin{axis}[
  grid style={line width=.1pt, draw=gray!10},
  major grid style={line width=.2pt,draw=gray!10},
  minor tick num=5,
  grid=both,
  ymode=log,
  xmode=log,
  title=\textbf{Runtime, strong scale},
  title style={yshift=-2.1ex,},
  legend style={nodes={scale=0.68, transform shape}},
  xlabel=\#Cores,
  ylabel=Time (sec),
   width=7cm,height=5cm]
\addplot [mark=*, line width=1pt, blue] table [y=LULESH, x=cpu]{data_for_plots/lulesh_strong_time_O0.dat};
\addlegendentry{LULESH};
\addplot [mark=square*, line width=1pt, red] table [y=Backus, x=cpu]{data_for_plots/backus_strong_time_O0.dat};
\addlegendentry{\backus{} - Lagrange};
\addplot [mark=triangle*, line width=1pt, green2] table [y=Backus, x=cpu]{data_for_plots/backus_strong_euler_time_O0.dat};
\addlegendentry{\backus{} - Euler};
\end{axis}
 \draw [dashed] (3.97,0) -- (3.97,2.33);
% \node at (3.5,6) {\large \textbf{Time vs \#Cores in Strong scale}};
\end{tikzpicture}
%%%%%%%%%%%%%%%%%%%%%%%%%%%%%% STRONG TIME
    %\includegraphics{}
  \subcaption{The execution times of \backus{} and \LULESH{} on the strong scaling test.}
  \label{fig:time_strong}
\end{subfigure}
% STRONG EFFICIENCY
\begin{subfigure}{0.4\textwidth}
\centering
\begin{tikzpicture}
\begin{axis}[
  grid style={line width=.1pt, draw=gray!10},
  major grid style={line width=.2pt,draw=gray!10},
  minor tick num=5,
  grid=both,
  xlabel=\#Cores,
  title=\textbf{Efficiency, strong scale},
  title style={yshift=-2.1ex,},
  legend style={nodes={scale=0.68, transform shape}},
    % title=\textbf{Strong scale efficiency},
  ylabel=Efficiency,
   width=7cm,height=5cm]]
%\title{Lagrange, Efficiency, Strong scale}
\addplot [mark=*, line width=1pt, blue] table [y=LULESH, x=cpu]{data_for_plots/lulesh_strong_O0.dat};
 \addlegendentry{LULESH}
\addplot [mark=square*, line width=1pt, red] table [y=Backus, x=cpu]{data_for_plots/backus_strong_O0.dat};
 \addlegendentry{\backus{} - Lagrange}
\addplot [mark=triangle*, line width=1pt, green2] table [y=Backus, x=cpu]{data_for_plots/backus_strong_euler_O0.dat};
 \addlegendentry{\backus{} - Euler}
\end{axis}
\draw [dashed] (1.43,0) -- (1.43,3.42);
% \node at (3.5,6) {\large \textbf{Efficiency vs \#Cores in Strong scale}};
\end{tikzpicture}
    \subcaption{The efficiencies of \backus{} and \LULESH{} on the strong scaling test.}
    \label{fig:eff_strong}
\end{subfigure}
% \vspace{0.3cm}
\begin{subfigure}{0.4\textwidth}
    \centering
\begin{tikzpicture}
\begin{axis}[
  grid style={line width=.1pt, draw=gray!10},
  major grid style={line width=.2pt,draw=gray!10},
  minor tick num=5,
  grid=both,
  title=\textbf{Runtime, weak scale},
  title style={yshift=-2.1ex,},
  legend style={nodes={scale=0.68, transform shape}},
   ymax=640,
%   ymode=log,
  xlabel=\#Cores,
   ylabel=Time (sec),
   width=7cm,height=5cm]]
\addplot [mark=*, line width=1pt, blue] table [y=LULESH, x=cpu]{data_for_plots/lulesh_weak_time_O0.dat};
 \addlegendentry{LULESH};
\addplot [mark=square*, line width=1pt, red] table [y=Backus, x=cpu]{data_for_plots/backus_weak_time_O0.dat};
 \addlegendentry{\backus{} - Lagrange};
\addplot [mark=triangle*, line width=1pt, green2] table [y=Backus, x=cpu]{data_for_plots/backus_weak_euler_time_O0.dat};
 \addlegendentry{\backus{} - Euler};
\end{axis}
% \node at (3.5,6) {\large \textbf{Time vs \#Cores in Weak scale}};
\end{tikzpicture}
    \caption{The execution times of \backus{} and \LULESH{} on the weak scaling test.}
    \label{fig:time_weak}
\end{subfigure}
\begin{subfigure}{0.4\textwidth}
    \centering
    \begin{tikzpicture}
\begin{axis}[
  grid style={line width=.1pt, draw=gray!10},
  major grid style={line width=.2pt,draw=gray!10},
  minor tick num=5,
  grid=both,
  xlabel=\#Cores,
  ylabel=Efficiency,
  title=\textbf{Efficiency, weak scale},
  title style={yshift=-2.1ex,},
  legend style={nodes={scale=0.68, transform shape}},
   width=7cm,height=5cm]
\addplot [mark=*, line width=1pt, blue] table [y=LULESH, x=cpu]{data_for_plots/lulesh_weak_O0.dat};
 \addlegendentry{LULESH}
\addplot [mark=square*, line width=1pt, red] table [y=Backus, x=cpu]{data_for_plots/backus_weak_O0.dat};
 \addlegendentry{\backus{} - Lagrange}
\addplot [mark=triangle*, line width=1pt, green2] table [y=Backus, x=cpu]{data_for_plots/backus_weak_euler_O0.dat};
 \addlegendentry{\backus{} - Euler}
\end{axis}
% \node at (3.5,6) {\large \textbf{Efficiency vs \#Cores in Weak scale}};
\end{tikzpicture}
    \caption{The efficiencies of \backus{} and \LULESH{} on the weak scaling test.}
    \label{fig:eff_weak}
\end{subfigure}
\label{fig:results}
\caption{\backus{} and \LULESH{} comparison on the strong and weak scaling test.}
\end{figure}
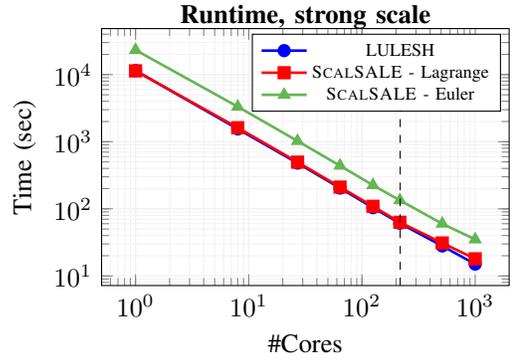
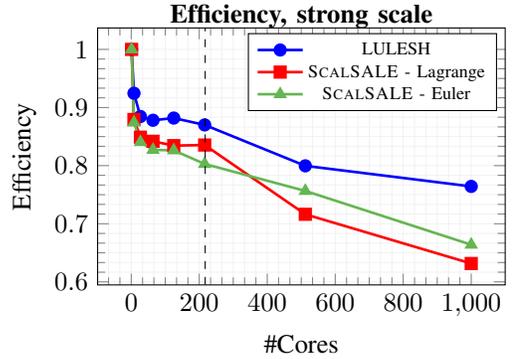
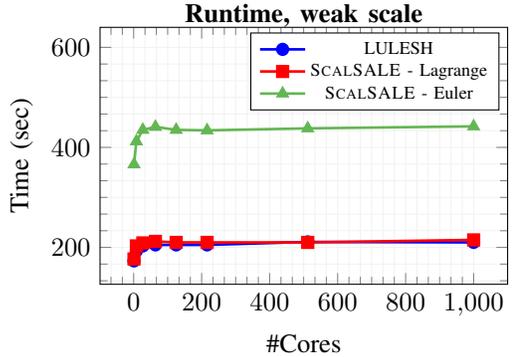
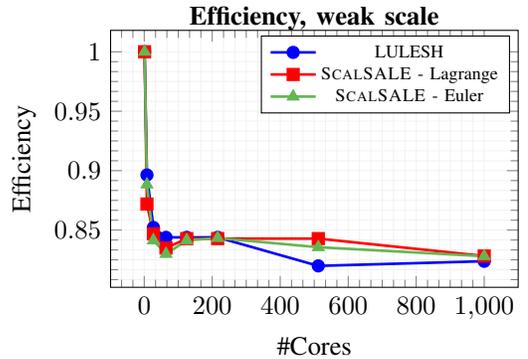
For the strong scale testing, a 3D Sedov-Taylor problem with a mesh size of \(480^3 = 110,592,000\) (480 cells in each axis) was defined. The total memory consumption of the problem is 80GB in \backus{} and 71GB in \LULESH{}.
%The full datafile description can be found on the GitHub page~\url{TODO}. 
\autoref{fig:time_strong} presents the total execution time of the strong scaling problem and \autoref{fig:eff_strong} presents the strong scaling efficiency. 

As expected, the execution time of \LULESH{} is the shortest as it is a highly optimized dedicated Sedov-Taylor hydrodynamic computation software. Nevertheless, the execution time of \backus{} with Lagrangian rezoning is comparable to \LULESH{}. 

Although Eulerian rezoning achieves the slowest execution time, it still presents excellent scaling capabilities even when normalized per cycle. Most importantly, Eulerian rezoning provides a broader range of hydrodynamic problems to be solved while maintaining the same speedup presented for the Sedov-Taylor problem.

A decrease in efficiency can be spotted at 1--8 cores due to the transition from a serial execution with no communication overhead to a distributed execution. Due to the nature of the parallelization scheme in 3D simulations, there is another decrease in efficiency at 8--27 cores, as 27 cores are the minimal number of processes to achieve the maximum amount of communication (exactly one process has 27 neighbors). Next, a slight decrease in efficiency can also be spotted at 27--64 cores for all three executions due to the introduction of a slightly slower medium that connects processes from different compute nodes. The final decrease in efficiency occurs at 216 cores (dashed line). For a fixed problem size, as the number of cores increases, the numerical workload of each MPI process decreases --- i.e., the number of cells per MPI process decreases. Although the message size also decreases (message size is of the size $O(N^2)$), it is less dramatic than the decrease in the numerical workload (of size $O(N^3)$). Hence, the communication overhead outweighs the numerical calculations, and the efficiencies of both benchmarks decrease. 

In addition, at 216--1,000 cores, due to the additional synchronization operations in \backus{}, the margin between the two codes increase as the workload decreases faster than the communication overhead. 
%The margin between the two can also be seen in strong scaling efficiency at 216--1,000 cores --- whereas \backus{} decreases more dramatically than \LULESH{} (as it performs more communications).

Nonetheless, \backus{} presents excellent strong scaling capabilities in both Lagrangian and Eulerian rezoning compared to \LULESH{}.

\subsection{Weak Scale}
For the weak scale comparison, the mesh of each MPI process was set to \(120^3 = 1,728,000\) (120 cells in each axis), similarly to~\cite{dosanjh2015re}. A mesh size of $120^3$ is equivalent to approximately 1.7GB in \backus{} and 1.1GB in \LULESH{}. As each computational node contains 32 cores, at full capacity, each node would consume up to 55GB in \backus{} and 35GB in \LULESH{}. \autoref{fig:time_weak} presents the total execution time of the weak scaling problem, and \autoref{fig:eff_weak} presents the weak scaling efficiency.

Similar to the strong scaling, \LULESH{} achieves the best execution time, while \backus{} with Lagrangian rezoning achieves a comparable execution time. However, contrary to strong scaling, each MPI process computes the same amount of numerical operations in weak scaling. Thus, the efficiency does not exhibit the a decrease in the efficiency. Nevertheless, the efficiency does tend to decrease slightly due to the additional cores that need to communicate with one another.

% Generally, weak scaling indicates the software's ability to be executed efficiently on many cores. As a scalable framework, the results indicate that \backus{} can manage an efficient large-scale execution. % 

Moreover, as the MPI implementation is generic and associated with the data structures and not the physical calculation, it has the potential to exhibit the same speedups on other problems in different domains. Thus achieving its targeted goal as a scalable benchmark framework.

% Despite performing more MPI operations, \backus{} still achieves comparable scaling capability to \LULESH{}. This might be due to the use of non-blocking communication, the MPI-3 collective operation --- \textit{MPI\_INeighbor\_alltoall} and the use of the topology features.
\section{Conclusion}
In this work, the problem with existing benchmarks was tackled by introducing a new scalable benchmark framework named \backus{}. The main objective of such a benchmark is to provide a more representative benchmark for scientific applications while maintaining simplicity. Contrary to existing benchmarks, \backus{} is flexible and designed such that it is easy to introduce new physical models, all while maintaining scalable speedups and low execution time. 

\backus{} is based on the well-known SALE scheme and is currently capable of solving the 1D, 2D, and 3D hydrodynamic equations in a structured grid with multi-material ALE capabilities (with an advection phase). A code with multi-material and ALE capabilities has the basic foundations for a code similar to hydrodynamic scientific applications.

\backus{} is written in Modern Fortran, utilizing object-oriented design patterns that provide the infrastructure to further develop the benchmark as a framework and the fast execution times of Fortran to support the code as optimized and scalable. Furthermore, Modern Fortran provides a well-balanced combination of fast execution times, especially for numerical computations, and sufficient support for object-oriented design patterns that allow easy modification and expansion of code. 

In order to further support the code as a framework, \backus{} is implemented with a two-layer design, the \textit{kernel} and the \textit{application}. The former handles the inherently parallel data structures (consequently, the physical quantities) that provide a transparent interface to the implementation of the distributed parallelization scheme. While the latter handles the numerical implementation and utilizes the parallel data structures to synchronize and keep up-to-date values. % Furthermore, as the MPI implementation is transparent, 

\backus{} parallelization scheme is based on the MPI-3 standard and utilizes two important features necessary for large-scale executions --- non-blocking collective communication and the processes/cores topology. Moreover, the MPI-3 standard provides wide support for hybrid parallelization schemes. Along with the separation of the numerical computations and the MPI implementation, MPI+X schemes can be added easily.

To evaluate \backus{} scaling capabilities, it was tested via the weak and strong scaling tests and compared to the well-known benchmark --- \LULESH{}. Despite performing more communication operations, \backus{} achieves excellent and comparable scaling results compared to \LULESH{} in both rezoning options. Moreover, as the parallelization scheme's implementation is independent of the numerical simulation, \backus{} can exhibit the same scaling capabilities for other physical domains. Hence, \backus{} achieves its targeted goal as a multi-material, multi-physical scalable benchmark framework.

\clearpage
\section*{Acknowledgment}
This work was supported by the Pazy foundation and the Lynn and William Frankel Center for Computer Science. Computational support was provided by the NegevHPC project~\cite{negevhpc}.

% \section*{References}

% Please number citations consecutively within brackets \cite{b1}. The 
% sentence punctuation follows the bracket \cite{b2}. Refer simply to the reference 
% number, as in \cite{b3}---do not use ``Ref. \cite{b3}'' or ``reference \cite{b3}'' except at 
% the beginning of a sentence: ``Reference \cite{b3} was the first $\ldots$''

% Number footnotes separately in superscripts. Place the actual footnote at 
% the bottom of the column in which it was cited. Do not put footnotes in the 
% abstract or reference list. Use letters for table footnotes.

% Unless there are six authors or more give all authors' names; do not use 
% ``et al.''. Papers that have not been published, even if they have been 
% submitted for publication, should be cited as ``unpublished'' \cite{b4}. Papers 
% that have been accepted for publication should be cited as ``in press'' \cite{b5}. 
% Capitalize only the first word in a paper title, except for proper nouns and 
% element symbols.

% For papers published in translation journals, please give the English 
% citation first, followed by the original foreign-language citation \cite{b6}.

\bibliographystyle{IEEEtran}
\bibliography{bibTex.bib}

\end{document}